\documentclass[twocolumn,prl,showpacs]{revtex4}

\usepackage{graphicx}
\usepackage{rotating}
\usepackage{amsmath}
\usepackage{amsfonts}
\usepackage{amssymb}
\usepackage{enumerate}
\usepackage{longtable}
\usepackage{dcolumn}
\usepackage{bm}

\begin{document}

\newcommand{\be}{\begin{equation}}
\newcommand{\ee}{\end{equation}}
\newcommand{\bn}{\begin{eqnarray}}
\newcommand{\en}{\end{eqnarray}}

\title{Fractionalized Fermi Liquid and $d$-wave Superconductivity in Ba$_{2}$CuO$_{3+x}$
}

\author{M. S. Laad$^{1}$, S. Koley$^{2}$ and A. Taraphder$^{3}$}

\affiliation{$^{1}$ The Institute of Mathematical Sciences, Taramani, Chennai 600113, India.\\
$^{2}$Department of Physics, Amity Institute of Applied Sciences, Amity University, Kolkata, 700135, India.\\
$^{3}$Department of Physics, Indian Institute of Technology Kharagpur, W.B. 721302, India
}
\date{\today}

\begin{abstract} Strongly enhanced superconductivity (SC) ($T_{c}=73$~K) in Ba$_{2}$CuO$_{3+x}$ (BCO) challenges the long-held paradigm of the irrelevance of the $d_{z^{2}}$-like states in cuprates. Using tight-binding band structure plus dynamical mean field theory (TB+DMFT), we find that a fractionalized Fermi liquid (FFL) is the normal state for the parameters realistic for BCO.  Transport properties exhibit strange metal behavior above $T_{coh}\sim 38$\,K, supporting a direct instability of this strange metal to SC.  Irrelevance of one-fermion inter-band mixing, itself a non-trivial consequence of the FFL state, preferentially enhances two-fermion hopping processes and provides a novel route for both SC and competing excitonic order(s) arising directly from the FFL. The proposed scenario also clarifies the role of $d_{z^{2}}$ band in shaping the physical responses of BCO.
\end{abstract}
\maketitle

\vspace{1.0cm}
  
{\bf Introduction}: Superconductivity (SC) in cuprates continues to spring surprises at regular intervals.  There is a general consensus ~\cite{PWAbook,ARPES-NatComm} that the physics of canonical cuprates is controlled by an {\it effective} one-band Hubbard model (1BHM) of planar Cu $3d_{x^{2}-y^{2}}-$O $2p_{\sigma}$ states, while  mixing~\cite{ref} with the $d_{z^{2}}$-orbitals is actually detrimental to SC.  This is facilitated by the elongated CuO$_{6}$ octahedron, pushing the $d_{z^{2}}$ orbitals well below the planar $d_{x^{2}-y^{2}}-p_{\sigma}$ states: only the latter cross the Fermi energy ($E_{F}$), in doped cuprates. 

Discovery of Ba$_{2}$CuO$_{3+x}$ (BCO) with sizably {\it compressed} CuO$_{6}$ octahedra~\cite{Uchida} overturns this paradigm.  Now, the $d_{z^{2}}-p_{z}$ states are 
pushed above the planar states, and both the $a=d_{x^{2}-y^{2}}$- and $b=d_{z^{2}}$-like bands cross $E_{F}$, with a type-II Dirac point about $0.3$~eV above $E_{F}$ as shown in the band structure calculations. These states are coupled by a $d$-wave form factor hybridization, $H_{hyb}=V_{ab}\sum_{k,\sigma}\gamma(k)(a_{k\sigma}^{\dag}b_{k\sigma}+h.c)$ with $\gamma(k)=$cos$k_{x}-$cos$k_{y}$.  This gives electron (at $\Gamma$) and hole (at $X$) pockets in the first Brillouin zone.  With both bands partially filled, one must deal with a two-band model~\cite{Scalapino}.  This raises several relevant issues: $(i)$ How distinct is BCO to the canonical cuprates?  $(ii)$ May ideas developed for canonical cuprates hold in BCO?  $(iii)$ Why is the SC $T_{c}=73$~K sizably higher than in isostructural hole-doped La$_{2-x}$Sr$_{x}CuO_{4}$ ($37$~K), where the $d_{z^{2}}$ band is irrelevant?
These issues are important in the quest for the enhancement of SC $T_{c}$ in cuprates, and suggest that optimizing the CuO$_{6}$ octahedral shape in cuprates holds a clue. Moreover, if the type-II Dirac point, only $0.3$~eV above $E_{F}$ in BCO, can be tuned to it, spin-orbit coupling- (SOC) preserving appropriate symmetries could open a gap and induce a topological Quantum Spin Hall phase~\cite{ARPES-NatComm}.  

\begin{figure}
\includegraphics[angle=270,width=\columnwidth]{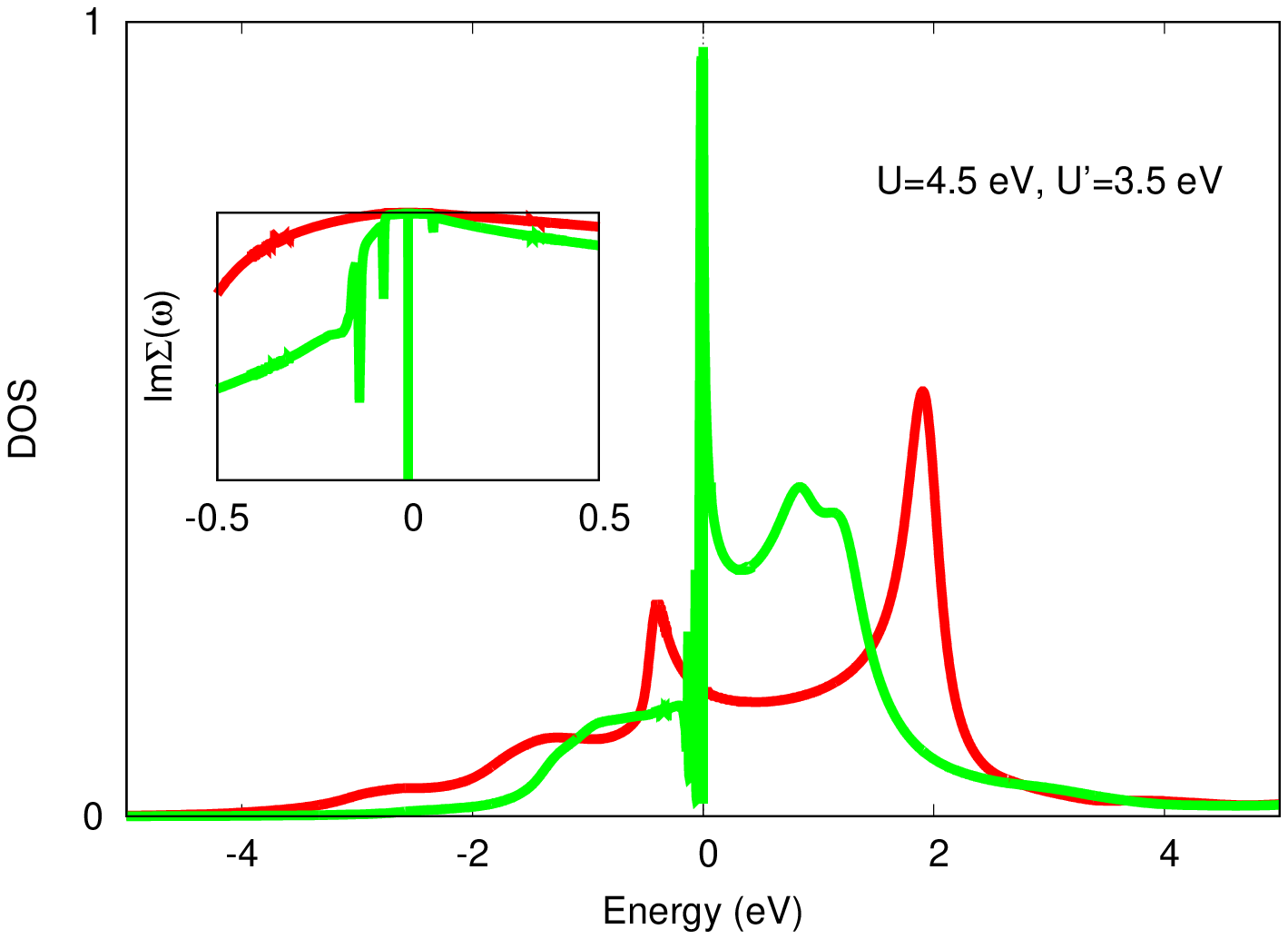}
	\caption{(Color Online) Orbital-selective DMFT spectra (red=d$_{x^{2}-y^{2}}$ and green=d$_{z^{2}}$) for BCO in main panel.  Inset shows the imaginary part of the self energy around $E_{F}=\omega = 0\text{ eV}$.  The pronounced pole in $\text{Im}\,\Sigma_{b}(\omega=0)$ shows selective localization of $b=d_{z^{2}}$ fermion states (see text).
	}
\label{fig1}
\end{figure}
Experimental data for BCO are scarce. Doped carriers~\cite{Uchida} reside in planar Cu $d_{x^{2}-y^{2}}-p_{\sigma}$ states, as in canonical cuprates.  But RIXS data~\cite{RIXS} show $D=1$ spin correlations and magnetic order involving Cu $d_{z^{2}}$ states, a feature specific to BCO.  The $D=1$ spin correlations survive in the well-doped case, putting a strong constraint on theory.  Though even pair-symmetry in BCO is unknown, this issue has been considered in weak~\cite{ref3} and strong~\cite{ref4} coupling views, with conflicting conclusions.  Since undoped BCO is an antiferromagnetic (AF) Mott insulator, we take the strong-coupling view and investigate the above issues $(i)-(iii)$ using DMFT in concert with tight-binding (TB)  band structure.  We find a novel fractionalized Fermi liquid (FFL) state, appearing as an orbital-selective Mott phase (OSMP: the $d_{z^{2}}$ states are Mott localized, while the planar $d_{x^{2}-y^{2}}$ states remain metallic.  Thus, we again end up with a 1BHM-like situation.  But in contrast to other cuprates, 
the reconstructed (by DMFT) lower Hubbard ($d_{z^{2}}$) band lies very close to $E_{F}$, and it cannot be integrated out to yield an 1BHM.  We describe its strange metal-like response, and detail how irrelevance of the interband one-fermion mixing in such an FFL phase simultaneously creates conditions necessary for a direct two-fermion instability of the FFL to SC.

\begin{figure}
(a)
\includegraphics[angle=270,width=0.8\columnwidth]{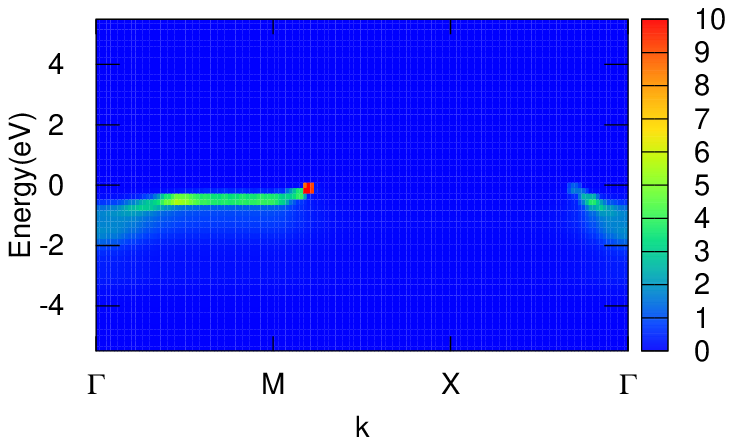}

(b)
\includegraphics[angle=270,width=0.8\columnwidth]{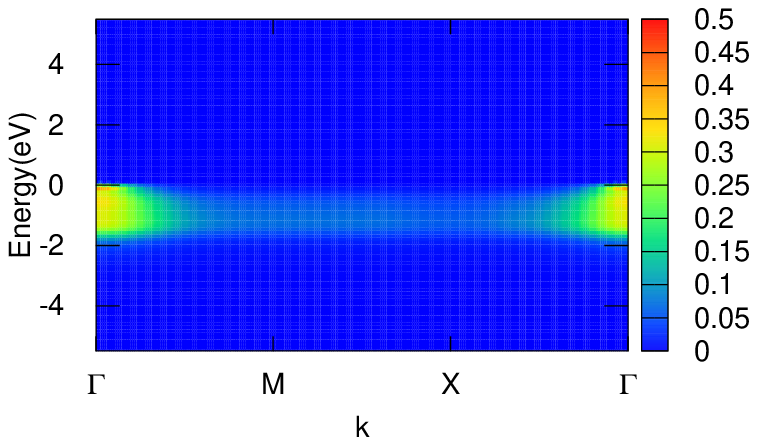}
\caption{(Color Online) Momentum-resolved ``ARPES'' spectral function 
	$A(\mathbf{k}, \omega)$ along high-symmetry points in the Brillouin zone 
	($\Gamma\text{–}\text{M}\text{–}\text{X}\text{–}\Gamma$) in 
	the OSMP: In (a), very low energy Landau quasiparticles co-exist with kink and smeared waterfall-like structures. In panel (b), 
	spectral weight vanishes at $E_{\text{F}}(=0)$ (see text), 
	resulting in an incoherent (lower) Hubbard band.} 
\label{fig2}
\end{figure}
We fit the bands of BCO at $x=0.25$ ($T_{c}=73$~K), with parameters used earlier~\cite{ref4}, shown in Fig.S1 of Supporting Material (SI1).  At this level, the Fermi surface (FS) of BCO comprises almost nested electron ($b=d_{z^{2}}$)) and hole $(a=d_{x^{2}-y^{2}}$) pockets.  In a weak coupling view, these are central to eventual instabilities of a Landau Fermi liquid (LFL) metal~\cite{chubukov}.  Such a scenario, however, is untenable in any OSMP. To study which applies to BCO, we solve a two-band Hubbard model with appropriate parameters, with MO-IPT as the impurity solver (see SI1).  DMFT drastically modifies DFT (TB) bands.  First, orbital-dependent Hartree shifts modify onsite energies that depend on the (selfconsistently determined) orbital occupancies: this renormalizes the interband charge transfer and hybridization.  More crucially, it non-trivially modifies the bands by capturing dynamical correlation-induced band narrowing via self-energies $\Sigma_{\mu}(\omega)$, ($\mu=a,b,$). At weak coupling, a Landau FL (LFL)  metal obtains at low energy.  With $U=4.5$~eV, $U'=3.5$~eV, and $J_{H}=0.5$~eV, we however find a OSMP~\cite{OSMP}, {\it i.e.}, a metal where nearly incoherent metallic $a$-band fermion quasiparticles (QPs) co-exist with Mott localized $b$-fermions.  This is in accord with earlier DMFT work~\cite{Held,chinese}.  Fig.~\ref{fig1} shows that Im$\Sigma_{b}(\omega=E_{F})$ exhibits a pole, while Im$\Sigma_{a}(\omega)\simeq -(\omega - E_{F})^{2}$ shows a correlated LFL behavior, but only below $|\omega-E_{F}|\leq 0.03$~eV. At larger energies, we recover the $-|\omega-E_{F}|$ behavior of a marginal-FL.  This is a fractionalized FL (FFL) metal~\cite{Sachdev2003} rather than a normal LFL: it is separated from the LFL by a topological transition that transmutes the zero in Im$\Sigma_{b}(\omega=E_{F})$ in the LFL phase into a pole
at $E_{F}$ in the FFL.  It hosts $b$-fermion local moments co-existing with metallic $a$-band QPs at low energy, and is an FS reconstruction, since the $b$ fermions do not anymore  contribute to the Fermi volume.  Moreover, Mott localized $b$-fermions can readily give the observed spin-$1/2$, $D=1$ Heisenberg spin chains via direct $b-b$ Anderson super-exchange along the CuO chain link in BCO: in the FFL, this co-exists with the $a$-band of nearly incoherent states.

{\bf Spectral And Transport Responses}:  Now, physical responses drastically deviate from those of a normal LFL.  Fig.~\ref{fig2} shows the ``ARPES'' band structure from DMFT, following an earlier recipe~\cite{Laad2015}.  An almost incoherent $\Sigma_{a}(\omega)$ and the pole in Im$\Sigma_{b}(\omega)$ at $E_{F}$ drastically modifies the correlated bands vis-a-vis the TB bands.  The LHB of $b$-fermions lies entirely below $E_{F}$: its weight is concentrated around the $\Gamma$ point and it is almost dispersionless.  Correlations dramatically reduce the $a$-band width, $W_{a}$, from $3.5$~eV to $~0.8-1.0$~eV.  Well-defined QPs are only found along $M-X$, and appreciable damping (from Im$\Sigma_{a,b}(\omega)$) strongly damps out states beyond $0.1-0.3$~eV.  A clear low-energy kink at $0.2$~eV is visible, as is a rapid change (blurred ``waterfall'') in the slope at $0.4$~eV close to the $\Gamma$ point (along $\Gamma -X$).  Previously attributed~\cite{Lebreuil2025,Uhrig} to coupling of fermions to a bosonic mode and to AF spin correlations, we find that local, dynamical charge and spin correlations in DMFT ({\it not} bosonlike at short length scales) can also produce them.  Finally, given an OSMP, any FS nesting-induced instabilities are untenable.

\begin{figure}
\includegraphics[angle=270,width=0.8\columnwidth]{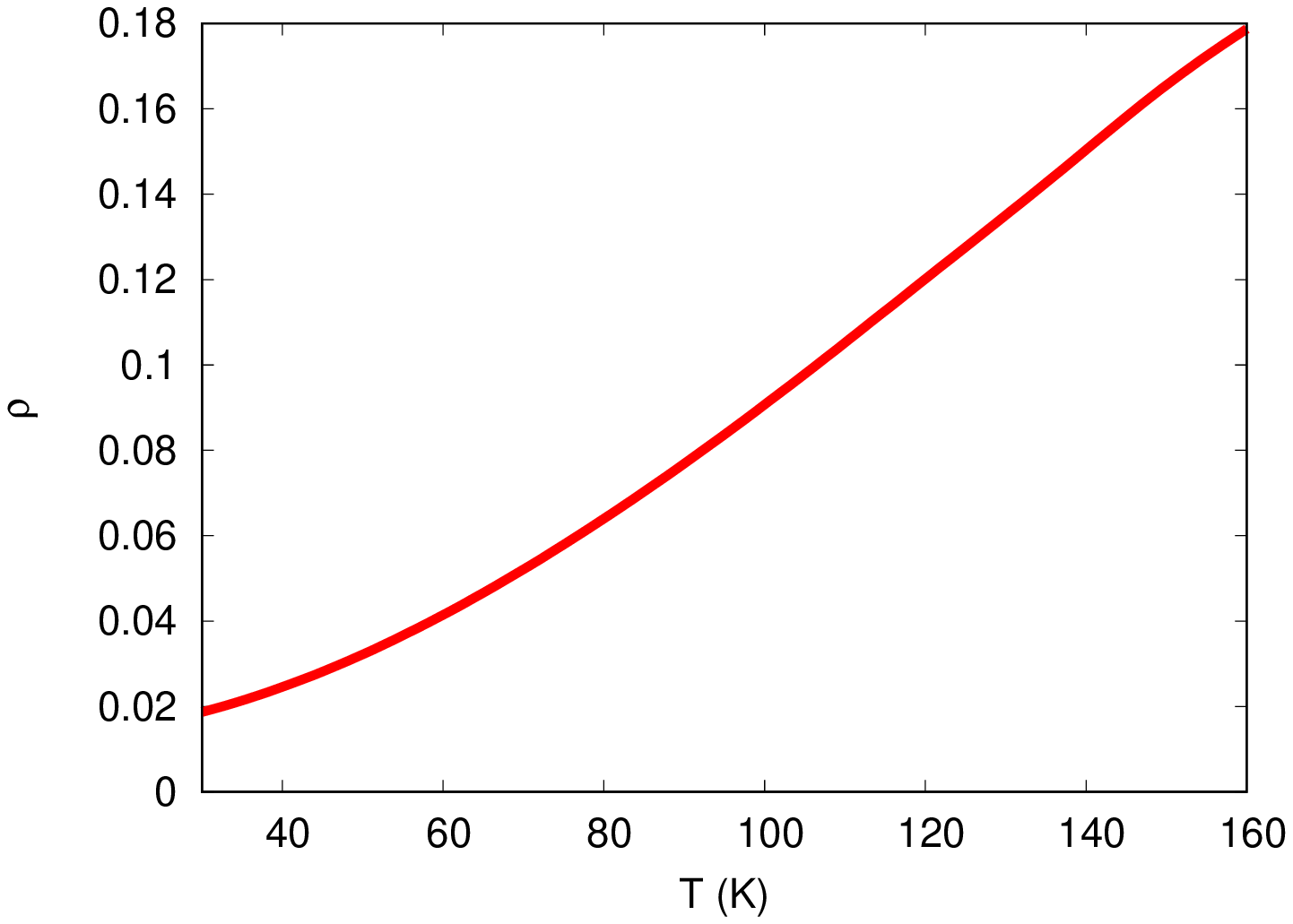}    
\caption{(Color Online) dc resistivity 
	$\rho_{\text{dc}}(T)$ (in arbitrary units) for 
	$\text{Ba}_2\text{CuO}_{3+x}$ ($T_{\text{c}} = 73\text{ K}$) shows a smooth crossover from $T$-linear behavior above $60\text{ K}$ to an apparent normal
	$\rho_{\text{dc}}(T)\propto T^2$) form below $30\text{ K}$.  Extrapolation of the $T$-linear part cuts the 
	$T$-axis at $T \approx 38\text{ K}$, marking a pseudogapped 
	metal coexisting with localized $b$-fermion moments.}
\label{fig3}
\end{figure}

\begin{figure}
(a)	
\includegraphics[angle=270,width=0.8\columnwidth]{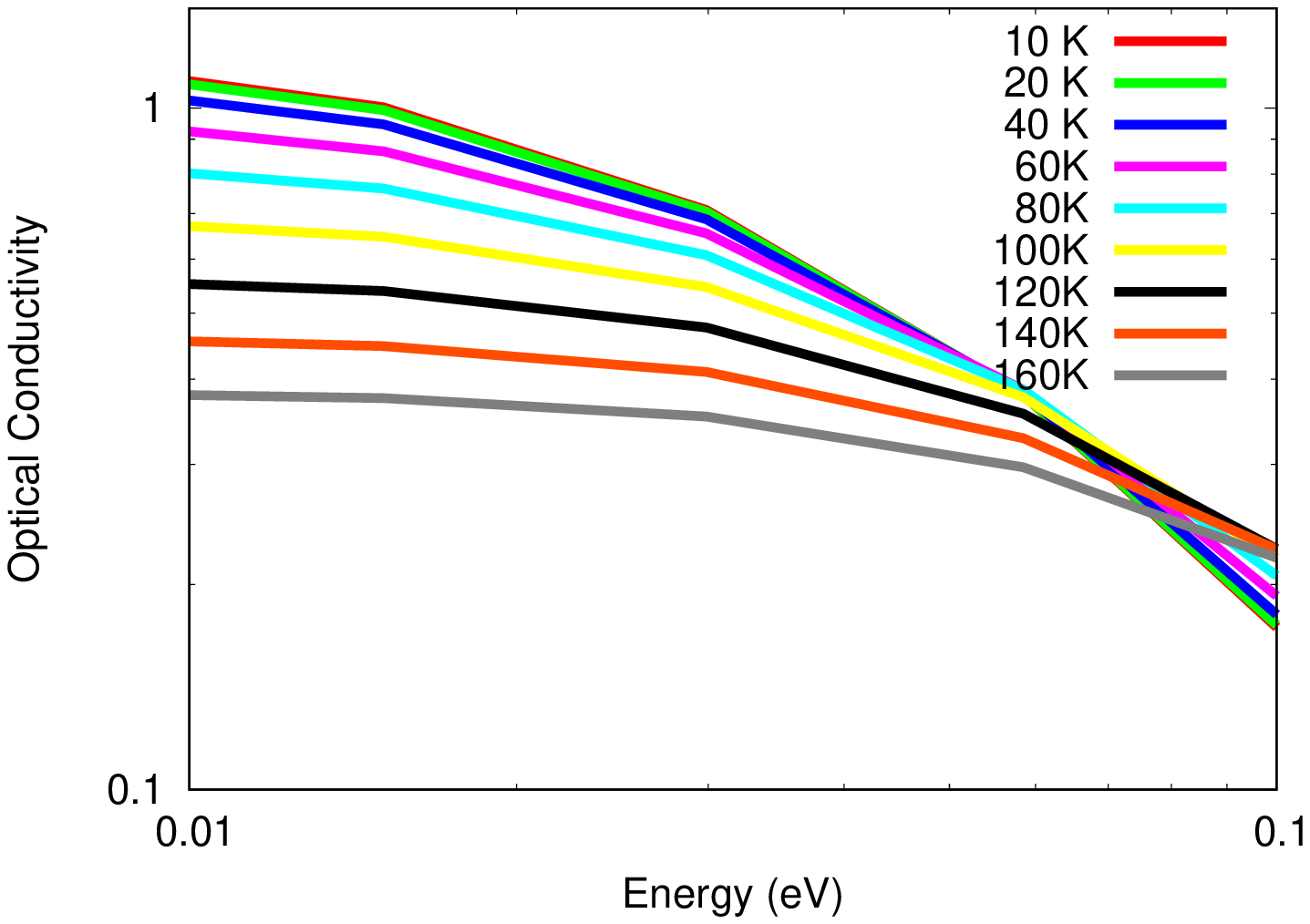}

(b)
\includegraphics[angle=270,width=0.8\columnwidth]{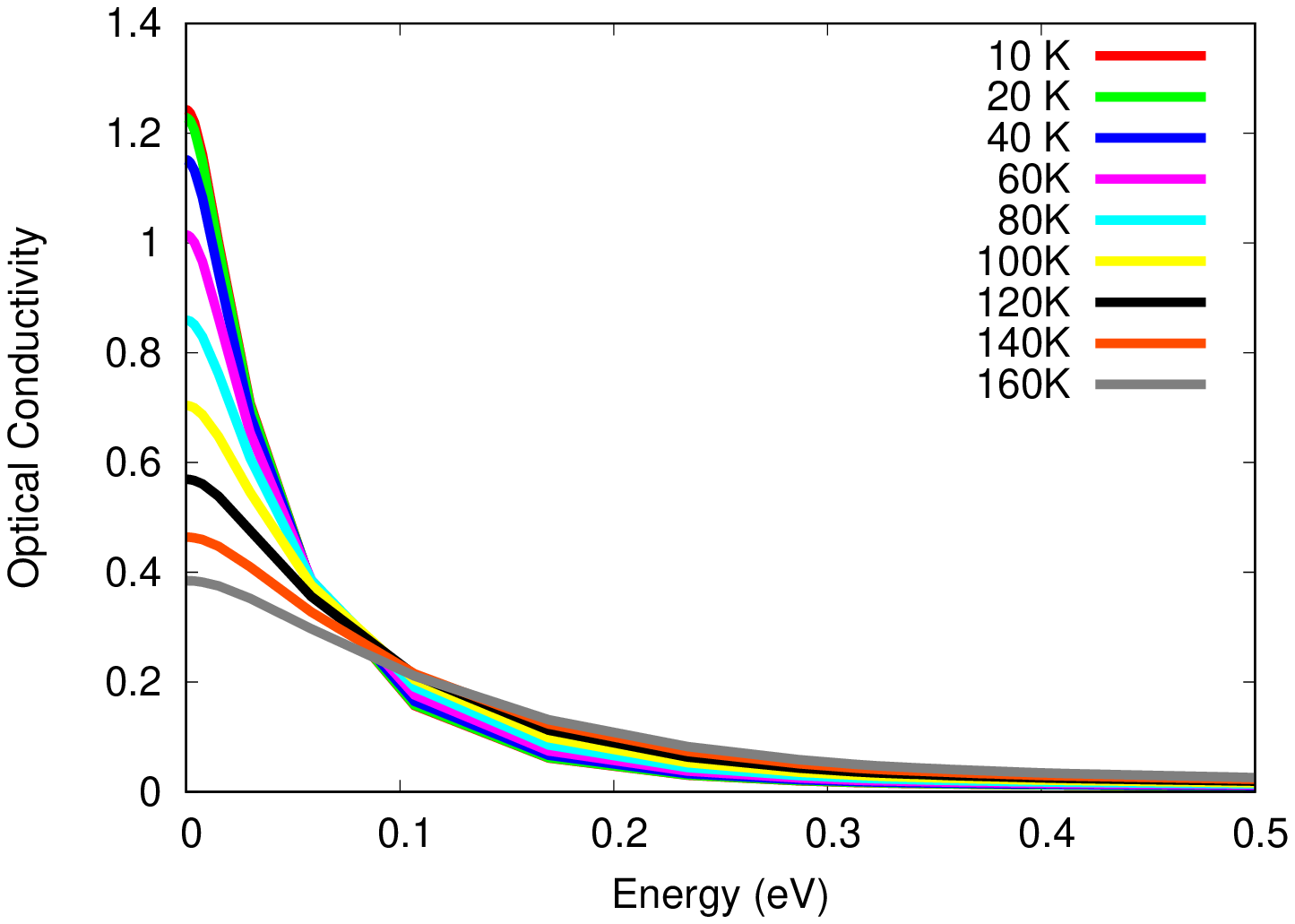}
\caption{(Color Online) Total low-energy intra-band optical conductivity 
	$\sigma_{xx}(\mathbf{q}=0, \omega; T)$ as a function of temperature 
	($10\text{–}160\text{ K}$): (a) log-log plot highlighting the 
	anomalous low-frequency power-law scaling regime for $0.03\leq\omega = 0.1
	\text{ eV}$, and (b) $\sigma_{xx}(\mathbf{q}=0,\omega; T)$ on a linear scale over 
	$0 \le \omega \le 0.5\text{ eV}$.} 
\label{fig4}
\end{figure}

Irrelevance of vertex corrections for the conductivities~\cite{Khurana, biermann} in DMFT simplifies their evaluation.  In Fig.~\ref{fig3}, we show the $dc$ resistivity for Ba$_{2}$CuO$_{3.25}$ (with $T_{c}=73$~K).  Notably, a wide linear-in-$T$ regime in $\rho_{dc}(T)$ smoothly evolves into an LFL-like form ($\rho_{dc}(T)\simeq T^{2}$) only below $30$~K.  Lest one conclude that this is an indication of an LFL metal, extrapolation of the $T$-linear component cuts the $T$-axis at $38$~K, indicating a pseudogapped (PG) metal.  Together with co-existent $b$-fermion local moments, this state is thus quite unconventional.  Moreover, $\rho_{dc}(T)\simeq T$ for $T>73$~K.  
The optical conductivity is computed using the Kubo formula with the DMFT Green functions as inputs.  In Fig.~\ref{fig4}, we show its low-energy part ($\sigma(\omega,T)$) up to $0.1$~eV as a function of $T$.  Remarkably, for $T\geq 38$~K, we find a non-QP, incoherent fractional power-law fall-off behavior, $\sigma_{xx}({\bf q=0},\omega)\simeq \omega^{-\alpha}$, in the range $0.03\leq\omega\leq 0.1$~eV, clearer in the inset.  The exponent, $\alpha$, weakly depends on $T$ and, for $T=80$~K, we estimate its average value, $\alpha\simeq 0.8$ in the range $0.03-0.1$~eV.  This non-Drude fall-off
marks the strange-metal-like optical response~\cite{PWAbook} in the $T$-regime where $\rho_{dc}(T)\simeq A\,T$.

{\bf Instability to Superconductivity}: If the pairing scale in BCO exceeds $0.03$~eV (see below), this
strange metal response will govern the eventual instability of the FFL metal to SC, with testable consequences~\cite{KotliarFeAs}.  The pair-glue function that enters the 
strong-coupling Eliashberg theory~\cite{PAllen} follows from the normal-state optical response as

\be
\alpha^{2}F(\omega)=\frac{\omega_{p}^{2}}{4\pi}\frac{d^{2}}{d\omega^{2}}[\omega \,Re(\frac{1}{\sigma(\omega)})]
\ee

\noindent yielding $\alpha^{2}F(\omega)\simeq \omega^{-(1-\alpha)}$.  Here, $\omega_{p}$ is the renormalized plasma frequency.  This infra-red singular $\gamma$-pairing case has been analyzed in detail~\cite{KotliarFeAs}.
For $\gamma=(1-\alpha)=0.2$, modest enhancement of the BCS gap-to-$T_{c}$ ratio ($y=2\Delta(0)/k_{B}T_{c}$) obtains: for $d$-wave pairing, the BCS value of $y$ is 4.2, so that  $y\simeq 4.3-4.5$.  Though modest, this should be taken cautiously: having an OSMP implicates strong phase fluctuations (PF) in the actual SC
instability.  Eliashberg theory cannot capture this effect.  The above mean-field estimate~\cite{KotliarFeAs} must thus be taken to be the ratio of the pair gap to the {\it pair formation} temperature, $T_{p}$, {\it i.e.}, to $y_{p}=2\Delta(0)/k_{B}T_{p}$.  Interestingly, our estimate is very close to the $y_{p}$ value for the pseudogap scale in canonical cuprates~\cite{Tacon}.  Since $T_{c}<T_{p}$ must hold due to enhanced PF, we expect a larger $y$ value in BCO.  This is testable, and would be of great interest.

What about microscopics of SC in BCO?  In usual cuprates, momentum selective Mottness requires appeal to CDMFT~\cite{Katsnelson}.  In BCO, the situation is even more complex, because presence of the $d_{z^{2}}$ Hubbard bands very close to $E_{F}$ prevents the option of integrating them out.  We ask, {\it what controls the effective pair potential that ultimately drives SC?}
To proceed, we first notice that {\it coherent} interband mixing is irrelevant in the OSMP (FFL) because transferring an $a$-fermion into the $b$-band (via $V_{ab}(k)$) is now dynamically blocked: since the $b$-states exist only as lower-(LHB) and upper Hubbard band (UHB) states, a low-energy state obtained by inserting an $a$-fermion into this sector must necessarily create a doublon, since it cannot occupy the (singly occupied) LHB of $b$-states.  Because this ``final state'' lies in the UHB in the $b$-sector, it has no overlap with any one-fermion state anymore.  It dynamically suppresses coherent $a-b$ mixing at one-fermion level, driving it irrelevant~\cite{Laad2026}. 

Given this irrelevance to $O(V_{ab})$, we appeal to an analogy with a similar situation known in the context of coupled $D=1$ Luttinger liquids (LL).  There, coherent {\it interchain} one-fermion hopping (the analogue of $V_{ab}(k)$ here) renormalizes to zero because of spin-charge separation.  The upshot~\cite{Kusmartsev} is that interchain two-fermion hopping processes in the particle-hole (ph) and particle-particle (pp) channels become more relevant (at second order in interchain hopping), leading to the direct instability of coupled LLs to either density-wave or SC order.  In our case, replacing each LL chain by a local site, an irrelevant
$V_{ab}(k)$ in the infra-red (FFL state) leads to a preferential relevance of two-fermion ph- and pp-hopping processes between neighboring {\it local sites to second order in $V_{ab}(k)$}.  To this order about the high-dimensional FFL fixed point, the effective two-fermion ``residual interaction'' is
\be
\begin{split}
	H_{res}\simeq -\frac{4V_{ab}^{2}}{U'}&\sum_{k,k',\sigma,\sigma'}\gamma(k)\gamma(k')(a_{k\sigma}^{\dag}b_{k\sigma}+h.c)\\
&(a_{k'\sigma'}^{\dag}b_{k'\sigma'}+h.c)
\end{split}
\ee  
\noindent $H_{res}$ contains terms $a_{k\sigma}^{\dag}a_{-k,-\sigma}^{\dag}b_{-k,-\sigma}b_{k\sigma}$ and $a_{k\sigma}^{\dag}b_{k\sigma}b_{k'\sigma'}^{\dag}a_{k'\sigma'}$.  The first is a Cooper pair hopping from the $a$-band to the localized $b$-sector and vice-versa, while the second hops an interband (singlet or triplet) exciton from $k$ to $k'$.  $H_{res}$ is an anisotropic Heisenberg pseudospin Hamiltonian (SI2), since the pair and exciton operators satisfy separate SU$(2)$ algebras. This can now be understood as a {\it charge Kondo-RKKY liquid} (CKL)~\cite{Taraphder-Coleman}, where pair hopping is enabled at $O(V_{ab}(k)^{2})$ via a dynamically generated ``RKKY'' interaction between pair-pseudospins.  An irrelevant $V_{ab}(k)$ in the OSMP pre-empts any pair-pseudospin quenching.  Thus, SC arises as a direct instability of the FFL, when the $XY$-part of this {\it charge} RKKY coupling becomes relevant.  We also find (see SI2) that a ferro-orbital-plus-bond-excitonic order is the leading competitor with SC in BCO.  Even with small spin-orbit interaction, a spin-triplet excitonic condensate with $\langle a_{k\sigma}^{\dag}b_{k,-\sigma}\rangle\neq 0$ will be admixed in this channel: it implies a ${\bf Q}=0$ magnetic order, with small moments pointing normal to the faces of each CuO$_{6}$ octahedron.  This is the circulating current (CC) phase~\cite{VarmaCC} 
$-$ while not relevant in canonical cuprates, it could appear to compete with SC in BCO.  

In our OSMP, the metallic $a$-QPs can also give a separate, {\it intra-band} BCS instability, triggered by another residual interaction, $H_{res}'
\simeq -\frac{4t_{a}^{2}}{U}\sum_{k,k'}\eta(k)\eta(k')a_{k\uparrow}^{\dag}a_{-k\downarrow}^{\dag}a_{-k'\downarrow}a_{k\uparrow}$ with $\eta(k)=$cos$k_{x}-$cos$k_{y}$, arising from intersite $a-a$ super-exchange as in a planar 1BHM~\cite{Katsnelson}.  Having two pairing channels could enhance SC (see SI2) via positive feedback: for example, $d$-wave SC can arise first in the $a$-sector via $H_{res}'$, and inter-band proximity can then switch on the second pairing channel via $H_{res}$, or vice-versa.  This could yield higher $T_{c}$ in BCO compared to isostructural LSCO (where $H_{res}$ is inoperative).  In SI2, we show that true SC involves an additional charge-RKKY interaction between $a$-fermion pair pseudospins, mediated by a new mechanism of exchange of two-spinon excitations of the spin-$1/2$ $D=1$ Heisenberg chains in BCO, and estimate the SC $T_{c}$ from these two channels.  Finally, analysis of the inter-layer hopping form-factors in the $a,b$-fermion sectors (SI4) shows that hopping of $a=d_{x^{2}-y^{2}}$-band pairs is significantly favored over that of the $b=d_{z^{2}}$-pairs, due to strong {\it formfactor mismatch} in the latter case.  This enables fully three-dimensional $d$-wave SC.         

In a weak coupling view, absence of an OSMP results in an LFL metal and, with nested $a,b$ FS pockets, SC can arise via inter-band pair-hopping~\cite{chubukov}.  This looks similar to what our $H_{res}$ gives, but the physics is qualitatively distinct: we require a normal state OSMP, whence no nesting can occur.  This could be tested by normal-state ARPES and Hall number measurements.   The strong-coupling $t-J$ model view is not inconsistent with ours if the doped carriers predominantly populate the $a$-band states.  XAS data~\cite{Uchida} support this assumption.  Our results qualitatively accord with earlier DMFT work.  However, these do not treat an instability to SC.  Moreover, in SI3, we detail how our OSMP also gives an excellent qualitative accord with perturbed-LL behavior seen in 
RIXS data~\cite{RIXS}.  Any non-OSMP must yield a $D=3$ LFL metal due to a relevant $V_{ab}(k)$, in conflict with RIXS data.  Also, absence of SC gap opening in the $D=1$ spin chain in RIXS disqualifies pairing arising in this sector.  Finally, large electronic specific heat at low $T<<T_{c}$~\cite{Uchida} naturally arises from nodal $d$-wave Bogoliubov quasiparticles {\it and} from the gapless spinons of the (co-existing) $D=1$ chains.  Thus, our proposal respects the constraints set by limited, extant data.  A $d$-wave SC co-existing with $D=1$ LL in the $b$-sector marks a {\it fractionalized} SC in BCO (SI2,3,4).

To conclude, we model BCO by a two-orbital Hubbard model in its sizable coupling limit using TB+DMFT.  Its normal state is a novel FFL with linear-in-$T$ resistivity and non-Drude power law optical conductivity: both are characteristic signatures of the strange metal~\cite{PWAbook}.  We detail how such an FFL is indispensable in preferentially favoring two-fermion hopping vis-a-vis (quenched in OSMP) one-fermion hybridization.  It is precisely this aspect that opens the door to a direct instability of the FFL to a multi-band SC with dominant $d_{x^{2}-y^{2}}$-wave pair symmetry under appropriate conditions.

\newpage
{\bf Supplementary Information
}

{\bf SI1: Tight Binding Band structure in BCO}

 To formulate an effective multi-orbital low-energy
model for the high-$T_{c}$ cuprate $\mathrm{Ba}_{2}\mathrm{CuO}_{3+x}$, we construct
a two-band tight-binding Hamiltonian capturing the low
energy physics derived from the strongly hybridizing
Cu-$3d$ manifold~\cite{s1Maier2019}. Because oxygen-vacancy ordering and
distortion in this system lift the conventional energy
structure, both the planar $d_{x^{2}-y^{2}}$ and axial $d_{z^{2}}$ orbitals are important
 in the vicinity of the chemical potential. Using orbital indices $l\in\{a,b\}$, the non-
interacting tight-binding Hamiltonian is given by:
\begin{equation}
H_{0}=\sum_{k,\sigma}\Psi_{k\sigma}^{\dagger}\begin{pmatrix}\bar{\epsilon}_{a}+\xi_{aa}(k) & \xi_{ab}(k)\\ \xi_{ba}(k) & \bar{\epsilon}_{b}+\xi_{bb}(k)\end{pmatrix}\Psi_{k\sigma}
\end{equation}
where the spinor is defined as $\Psi_{k\sigma}=(c_{ak\sigma},c_{bk\sigma})^{T}$. The
on-site orbital energies are set to $\epsilon_{a}=0.624$~eV and
$\epsilon_{b}=-0.185$~eV, with the inter-orbital energy offset van-
ishing by symmetry $(\epsilon_{ab}=0)$. Taking the square-lattice
coordinate system where the intra-cell Cu-O bonds align
with the axes, the intra-orbital dispersions $\xi_{ll}(k)$ include
hoppings up to third-nearest neighbors:
\begin{align}
\xi_{ll}(k) ={}& 2t_{ll}(\cos k_{x}+\cos k_{y})+4t_{ll}^{\prime}\cos k_{x}\cos k_{y} \nonumber\\
&+2t_{ll}^{\prime\prime}(\cos 2k_{x}+\cos 2k_{y})
\end{align}
with intra-orbital hopping parameters for the $d_{x^{2}-y^{2}}$
band given by $(t_{aa},t_{aa}^{\prime},t_{aa}^{\prime\prime})=(-0.495,0.057,-0.099)$~eV, and for the $d_{z^{2}}$ band
by $(t_{bb},t_{bb}^{\prime},t_{bb}^{\prime\prime})=(-0.205,-0.016,-0.039)$~eV. The
symmetry-mandated odd parity under fourfold rotation
governs the inter-orbital hybridization $\xi_{ab}(k)=\xi_{ba}(k)$:
\begin{equation}
\xi_{ab}(k)=2t_{ab}(\cos k_{x}-\cos k_{y})+2t_{ab}^{\prime\prime}(\cos 2k_{x}-\cos 2k_{y})
\end{equation}
with inter-orbital transfer integrals $t_{ab}=0.318$~eV, $t_{ab}^{\prime}=0$, and $t_{ab}^{\prime\prime}=0.067$~eV~\cite{s1Maier2019}. 

\begin{figure*}
(a)
\includegraphics[angle=270,width=0.8\columnwidth]{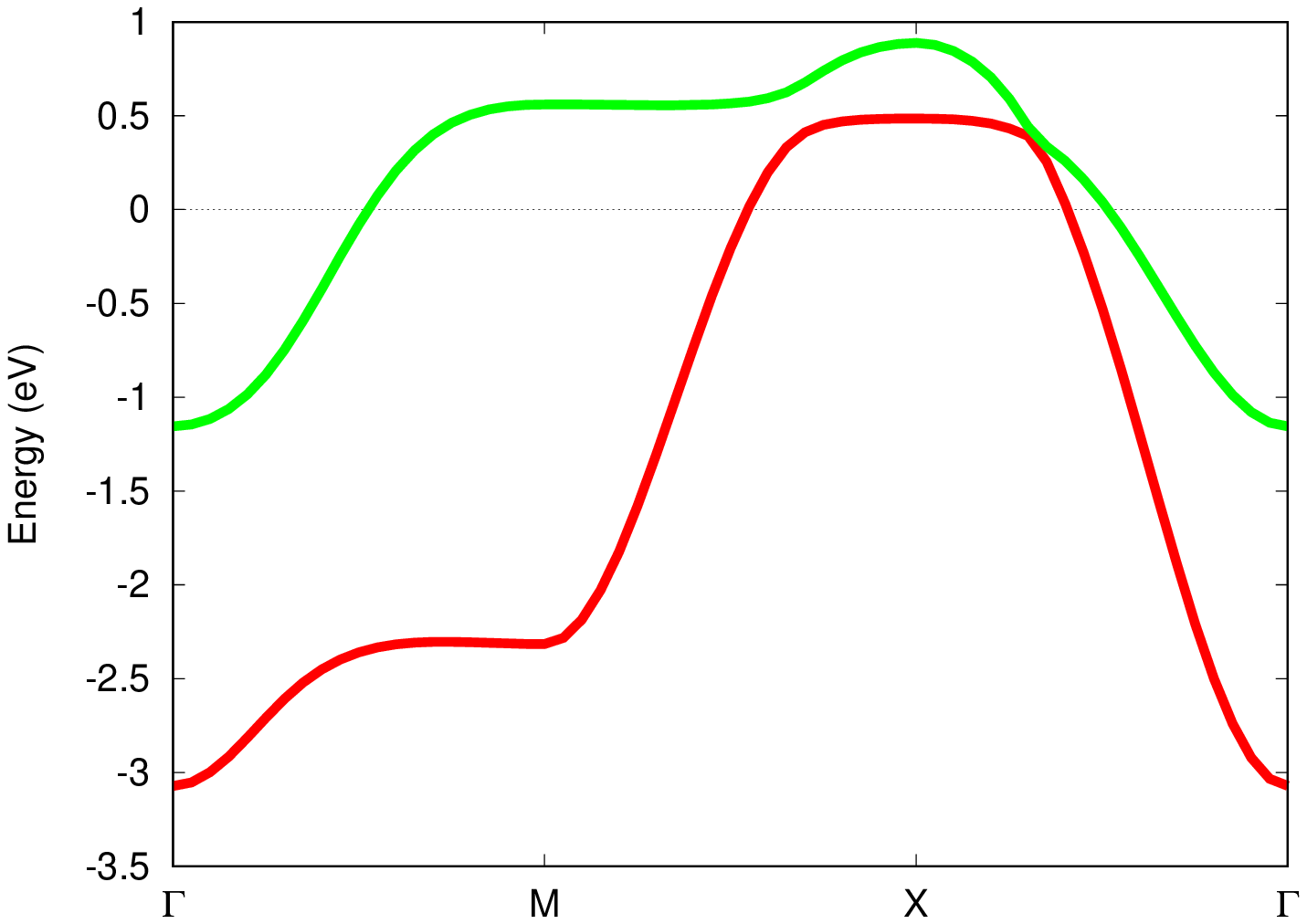}
(b)
\includegraphics[angle=270,width=0.8\columnwidth]{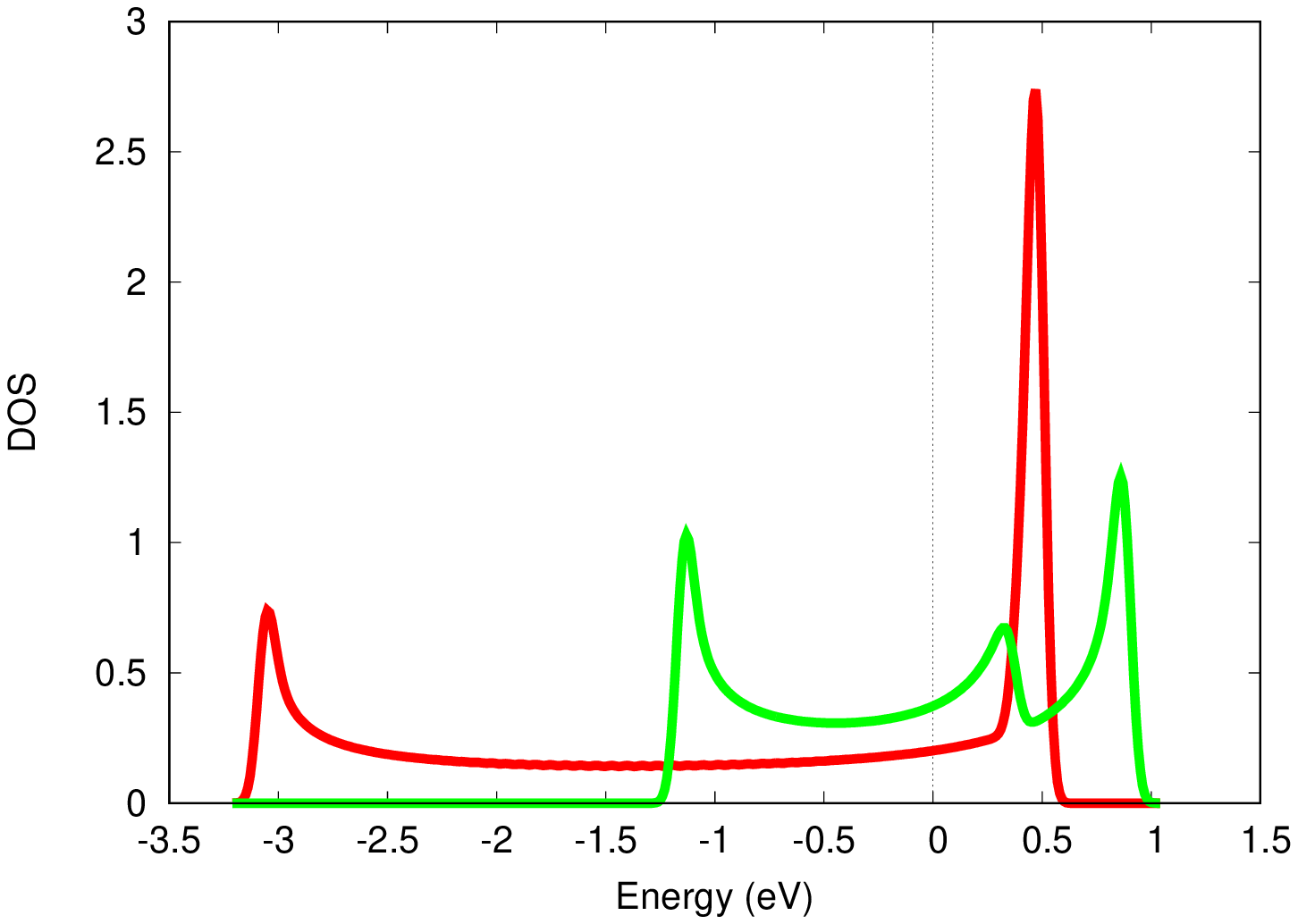}

(c)
\includegraphics[angle=270,width=0.8\columnwidth]{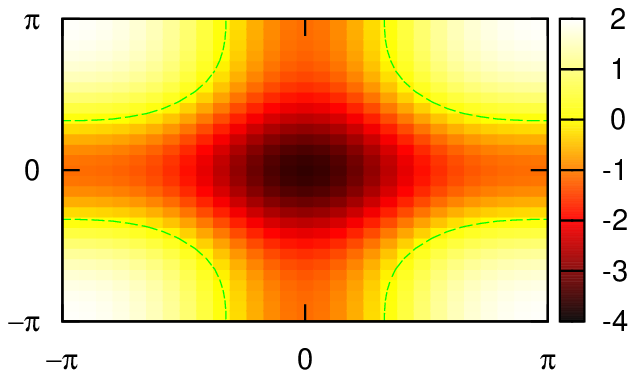}
(d)
\includegraphics[angle=270,width=0.8\columnwidth]{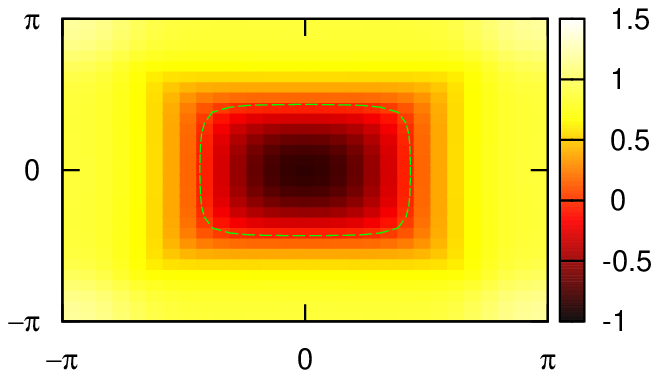}

        \caption{(Color Online) Non-interacting tight-binding electronic structure of the effective two-band model: (a) bare energy dispersion along high-symmetry paths in the two-dimensional Brillouin zone ($\Gamma\text{–}\text{M}\text{–}\text{X}\text{–}\Gamma$), with the Fermi level set to zero ($E_{\text{F}} = 0\text{ eV}$, horizontal dotted line); (b) corresponding bare orbital-resolved density of states (DOS), showing prominent Van Hove singularities near the band edges and close to $E_{\text{F}}$ (vertical dotted line).  (c),(d) show the Fermi pockets in BCO: the $b=d_{z^{2}}$-like band shows clear one-dimensional character, while the $a=d_{x^{2}-y^{2}}$-band gives four hole pockets, centered about $(\pm\pi,\pm\pi)$ in the first Brillouin zone.}
\label{figS1}
\end{figure*}

We show the tight-binding bands, the unperturbed local DOS, and the unperturbed Fermi pockets for the $a$- and $b$-fermion states in Fig.~\ref{figS1}.  Clearly, the $b$-band is much closer to being half-filled than the $a$-band, which corresponds to an ``overdoped'' band.  In this situation, we expect multi-orbital correlations to preferentially Mott
localize the $b$-band states, leaving the $a$-fermion states metallic.  But this is an OSMP, and gives a FFL state, precisely as described in the main text.
 
Moreover, in the main text, we have also employed the $c$-axis hopping form factors (FF) to derive the form factors for the hopping of $a$ and $b$-orbital Cooper pairs between neighboring layers.  In contrast to the canonical cuprates, the $b=d_{z^{2}}$-orbitals are active in BCO.  The one-fermion hopping form factor for $b$-orbitals in BCO is
constrained by the lattice structure, and reads

\be
H_{\perp}^{(b)}=t_{\perp}\sum_{k,\mu,\mu',\sigma}\beta(k)(b_{k,\mu,\sigma}^{\dag}b_{k,\mu',\sigma}+h.c)
\ee
where $t_{\perp}$ is the interlayer hopping integral for $b$-orbitals and $\mu\neq \mu'$ label neighboring layers.  In contrast to usual cuprates, this is sizable, $O(0.1t_{bb})$, in BCO.  The FF $\beta(k)$ is $\beta(k)=$cos$(k_{x}/2)$cos$(k_{y}/2)$cos$(k_{z}/2)$: it does not vanish along the nodal direction, so there is no ``nodal protection'' in this sector.

But the interlayer hopping FF for hopping of $a$-fermions between layers is very instructive.  We have

\be
H_{\perp}'=t_{\perp}'\sum_{k,\mu,\mu',\sigma}\lambda(k)(a_{k,\mu,\sigma}^{\dag}a_{k,\mu',\sigma}+h.c)
\ee
where $t_{\perp}'$ is the interlayer hopping integral for $a$-orbitals.  The FF $\lambda(k)$ is given by $\lambda(k)=($cos$k_{x}-$cos$k_{y})^{2}$cos$(k_{z}/a)$.  This exhibits
``nodal protection'', but with an additional cos$(k_{z}/2)$ modulation along the $c$-axis.  This FF arises from the fact that an $a$-fermion can only hop between neighboring layers by first mixing with a $b$-state via $V_{ab}(k)$, then hopping the $b$-fermion between layers, and finally by mixing it with an $a$-fermion state, again via $V_{ab}(k)$.
Being a higher-order process (by $O(V_{ab}^{2})$), we expect $t_{\perp}'$ to be smaller than $t_{\perp}$.

One might inquire about the fate of our results if we had included the $H_{\perp}, H_{\perp}'$ into the tight-binding band structure used in the main text.  While this would modify the $D=2$ FS pockets to $D=3$, the main effect on the unperturbed LDOS will be to slightly broaden the $b$-fermion band width, leaving the $a$-fermion band width almost unchanged.  Moreover, the $b$-fermion level will still lie above that of the $a$-fermions in BCO, and the $b$-band will still be much closer to half-filling than the $a$-band.  Thus, we expect that the OSMP we found in the main text will again occur in this case for slightly different values of the Hubbard parameters $U,U'$. 

\vspace{0.5cm}

{\it Multi orbital DMFT}:To treat dynamical correlation ef-
fects and multi-orbital electronic interactions beyond the
static mean-field level, we include local Coulomb terms:
\begin{equation}
H=H_{0}+\sum_{i,l}U_{l}n_{il\uparrow}n_{il\downarrow}+\sum_{i,\sigma,\sigma^{\prime}}\left(U^{\prime}-\frac{1}{2}J_{H}\delta_{\sigma\sigma^{\prime}}\right)n_{i1\sigma}n_{i2\sigma^{\prime}}
\end{equation}
where $U_{l}$ represents the intra-orbital Coulomb interac-
tion, $U^{\prime}$ the inter-orbital density-density repulsion, and
$J_{H}$ the Hund's coupling. We solve this interacting lat-
tice Hamiltonian within the framework of multi-orbital
Dynamical Mean-Field Theory (MO-DMFT)~\cite{s3Kotliar2006,s4Gull2011}. The lattice
problem is mapped onto an effective self-consistent multi-
orbital Anderson impurity model, where the local matrix
Green's function $\hat{G}_{\mathrm{loc}}(i\omega_{n})$ is obtained via the Brillouin-
zone integration:
\begin{equation}
\hat{G}_{\mathrm{loc}}(i\omega_{n})=\int_{\mathrm{BZ}}\frac{d^{2}k}{(2\pi)^{2}}[(i\omega_{n}+\mu)\hat{\mathbb{I}}-\hat{H}_{0}(k)-\hat{\Sigma}(i\omega_{n})]^{-1}
\end{equation}
Here, $\mu$ is the chemical potential tuned self-consistently
to establish the filling associated with $x=0.25$,
and $\hat{\Sigma}(i\omega_{n})$ is the momentum-independent local self-energy matrix. 
The dynamical Weiss host $\hat{G}_{0}^{-1}(i\omega_{n})=\hat{G}_{\mathrm{loc}}^{-1}(i\omega_{n})+\hat{\Sigma}(i\omega_{n})$ specifies the effective impurity 
environment. To compute the dynamic self-energy $\hat{\Sigma}(\omega)$ directly on the real-frequency axis at finite temperatures,
we employ the multi-orbital Iterated Perturbation The-
ory (MO-IPT) as the impurity solver~\cite{s5Laad2003,s6Koley2015}.  MO-IPT constructs the local self-energy interpolating between the exact weak-coupling
(second-order perturbation in $U,U'$) and the atomic limits:

\begin{equation}
\Sigma_{ll^{\prime}}^{(2)}(\tau)=\sum_{m,n,p}\mathcal{U}_{lmnp}\mathcal{G}_{0,mp}(\tau)\mathcal{G}_{0,nl^{\prime}}(-\tau)\mathcal{G}_{0,pn}(\tau)
\end{equation}

where $\mathcal{U}$ captures the tensor of local intra- and inter-
orbital Coulomb vertices, followed by standard IPT cor-
rections to yield the correct high- and low-frequency
asymptotic behaviors at arbitrary filling~\cite{s5Laad2003,s6Koley2015}. The converged
DMFT solutions provide the matrix spectral functions
$\hat{A}(k,\omega)=-\frac{1}{\pi}\mathrm{Im}\,\hat{G}(k,\omega+i0^{+})$ detailing orbital-selective
spectral weight transfers, quasiparticle lifetimes, and
interaction-induced band renormalizations.

\vspace{0.5cm}

{\bf SI2: Charge-Kondo-RKKY Model}  Here, we detail how a charge-Kondo-RKKY like model of pair-pseudospins arises from the structure of the residual interactions in the
main text.  We begin with $H_{res}$ in the main text, given by

\begin{equation}
\begin{split}
H_{\text{res}} \simeq {} & \frac{+4V_{ab}^{2}}{U'} \sum_{k,k',\sigma,\sigma'} \gamma(k)\gamma(k') (a_{k\sigma}^{\dag}b_{k\sigma} + \text{h.c.}) \\
& (a_{k'\sigma'}^{\dag}b_{k'\sigma'} + \text{h.c.})
\end{split}
\end{equation}
containing terms like 

\be
(\gamma(k)a_{k\sigma}^{\dag}a_{-k,-\sigma}^{\dag})(\gamma(k)b_{-k,-\sigma}b_{k\sigma})
\ee
 and 
 
 \be
 (\gamma(k)a_{k\sigma}^{\dag}b_{k\sigma})(\gamma(k')b_{k'\sigma'}^{\dag}a_{k'\sigma'})
 \ee
   With $\gamma(k)=$cos$k_{x}-$cos$k_{y}$, the first term hops an intra $a$-orbital $d$-wave pair onto a $b$-orbital and vice-versa,
while the second hops a $d$-wave singlet $(\sigma'=\sigma)$ or triplet $(\sigma'=-\sigma)$ exciton from $k$ to $k'$.  Clearly, pair pseudospins, constructed as 
$P_{k,-k,a}^{+}=a_{k\sigma}^{\dag}a_{-k,-\sigma}^{\dag}$, $P_{-k,k,a}^{-}=a_{-k,-\sigma}a_{k\sigma}$ and $P_{k,-k,a}^{z}=(1-n_{a,k,\sigma}-n_{a,-k,-\sigma})/2$ (and the $b$ pseudospins defined the same way) satisfy the usual (pseudo)spin-$1/2$ SU$(2)$ algebra.  The exciton pseudospins satisfy a separate SU$(2)$ algebra.  Focussing on the pairing sector, we can recast $H_{res}$ as follows.

\be
H_{res}=J_{ab}\sum_{k,k'}\gamma(k)\gamma(-k)(P_{k,-k,a}^{+}P_{-k,k,b}^{-}+h.c)
\ee
where $J_{ab}=(4V_{ab}^{2}/U')$.  We notice that the linear-in-$V_{ab}$ term, $V_{ab}(k)(a_{k\sigma}^{\dag}b_{k\sigma}+h.c)$, is irrelevant in the OSMP or FFL state,as discussed in detail in the main text.  This leads to irrelevance of any Kondo effect at this order, and the pair pseudospins are always unquenched.  Thus, no correlated charge-Kondo Fermi liquid state can occur, but a pair {\it liquid} state with strong ``preformed'' and incoherent pair-pseudospin fluctuations can still obtain.  Hence, the FFL normal state now undergoes
a direct instability to a paired state, described by coherent transfer of $d$-wave pairs between $a,b$ orbitals.  This is not yet a global SC, which requires an effective attractive interaction between $a$-pairs (see below for a specific mechanism, tailored to the OSMP in BCO).   It is instructive to observe that $H_{res}$ elevates the local gauge invariance that obtains in the OSMP, due to the relation $[n_{i,b},H]=0$ for all $i$ ($i$ is the site index for the original two-orbital Hubbard model), to a global $U(1)$ 
invariance.  Without $H_{res}$, Elitzur's theorem precludes any SC at any finite $T$.  Residual interactions evade this constraint, and give rise to a coherent inter-band pair hopping.

In BCO, there is another, {\it intraband} pair channel, arising from the metallic $a$-fermion sector in the OSMP.  By itself, it can lead to $d$-wave SC in the same way as it does for the usual one-band Hubbard model (1BHM)  in $D=2$.  With two pairing channels, positive feedback between them can enhance $d$-wave SC over and above that set in a 1BHM: we describe how this can occur below in this section.  
For example, $d$-SC can emerge first on the intraband sector, and the $a$-fermion pairs can delocalize via two distinct channels: via $H_{res'}$ involving the same mechanism as in a 1BHM, or via $H_{res}$, involving the interband channel discussed above.  The reverse possibility can also be operative.  This enhanced possibility for $d$-wave pairs to delocalize can push up the SC $T_{c}$.  

We mentioned above that true SC requires an effective attractive potential between $a$-fermions.  This attractive interaction has to be generated in the OSMP we find in the main text, where $b$-ferrmions give rise to a $D=1$ spin-$1/2$ chain, co-existing with nearly incoherent $a$-fermions.  This will also allow us to estimate the order-of-magnitude of the SC $T_{c}$ from each of the two channels as above.  We note that this is {\it not} exact, but an order-of-magnitude estimate: it gives insight into how presence of two channels can boost $d$-wave SC in BCO.  We begin by noting that the first channel, $H_{res}$, delocalizes $a$-band pairs by hopping them into the $b$-band: but the latter is a $D=1$ spin-$1/2$ Heisenberg spin chain (see SI3).  It's elementary excitations are spinon pairs, carrying spin but no charge.  Thus, the $a$-band pair can only transfer a pair of spinons into the $D=1$ spinon Fermi sea, which can readily absorb them, since the spinon Fermi sea is gapless.  This leaves behind a spinless holon pair in the $a$-sector.  This holon pair can subsequently re-absorb a spinon pair from the spinon Fermi sea, and hop off back into the $a$-band as a usual Cooper pair.
This is a process tailored to the specific situation of an OSMP in BCO, and is quite unconventional.  It is reminiscent of Anderson's RVB~\cite{s2PWAbook} ideas.  Specifically, the $a$-fermion pairs delocalize by exchanging (emitting and absorbing) charge-neutral spinon pairs from the $D=1$ Heisenberg spin chain.  It is thus a higher (fourth-order) order process in $V_{ab}(k)$, or second order in $J_{ab}$.  Explicitly, we may write

\be
H_{res}^{(aa)}\simeq J_{ab}^{2}\sum_{k,\sigma}\chi_{s\bar{s}}(2k_{F})\gamma^{2}
(k)a_{k,\sigma}^{\dag}a_{-k,-\sigma}^{\dag}a_{-k,-\sigma}a_{k,\sigma}
\ee
with $\chi_{s\bar{s}}(k-k'=2k_{F})$ (since $k'=-k=k_{F}$ for SC) the spinon pair-susceptibility entering the pair kernel.  Approximating it by $\rho_{s}(0)$, the spinon-pair density-of-states at $2k_{F}$, we estimate the latter to be $\rho_{s}(0)\simeq 1/W_{s}$, the spinon band width.  The coupling constant entering $H_{res}^{(aa)}$ is then $O(J_{ab}^{2}/W_{s})$.  Using $J_{ab}\simeq 0.05-0.1$~eV and $W_{s}=4J_{c}\simeq O(1)$~eV, we estimate $T_{c}\simeq J_{ab}^{2}/W_{s}=34.8$~K.  In BCO, however, there is a second, intra-band pair channel in the $a$-fermion sector.  With suitable choice of parameters, this can separately yield $T_{c}\simeq O(30-40)$~K.  Though, of course, the full $T_{c}$ is not simply expressible in terms of these two contributions, we propose that positive feedback between these two pairing channels can enhance $T_{c}$ in BCO relative to isostructural LSCO, where $H_{res}$ is irrelevant. 

Notice that our unconventional mechanism is reminiscent of a ``RKKY'' coupling.  However, this RKKY-like coupling is between $a$-band {\it charge} pseudospins, mediated by {\it spinon pairs}.  It naturally yields a $d_{x^{2}-y^{2}}$-wave SC in the $a$-band sector co-existing with a $D=1$ spin-$1/2$ Heisenberg chain the the $b$-fermion sector.  We thus obtain a {\it fractionalized} SC as a novel consequence of an underlying normal state FFL found in the main text.      

\vspace{0.5cm}

{\it Competing Order in BCO}:  But there is also a competing interaction in $H_{res}$ above, given by the inter-band exciton part:

\be
H_{res}^{(exc)}\simeq J_{ab}\sum_{<i,j>,\sigma,\sigma'}(a_{i\sigma}^{\dag}b_{j\sigma})(b_{j\sigma'}^{\dag}a_{i\sigma'})
\ee
which is expandable as a sum of a nearest neighbor interband interaction, $-J_{ab}\sum_{\sigma}[(1-n_{ia\sigma})(1-n_{jb\sigma})+(1-n_{jb\sigma})]$ for $\sigma'=\sigma$, and to an interband transverse spin fluctuation term, $-J_{ab}(S_{ia}^{+}S_{jb}^{-}+h.c)$, for $\sigma'=-\sigma$.  The first can be re-expressed as
\be
H_{RKKY}^{zz}=-J_{ab}\sum_{<i,j>}[4P_{i,a}^{z}P_{j,b}^{z}+(1-n_{jb\sigma})]
\ee
which adds a longitudinal term to the pseudospin ``RKKY'' model.  It also renormalizes the $b$-band energy level, shifting it further upward by $J_{ab}$.  We thus arrive at a {\it charge}-pseudospin $P=1/2$ XXZ model.  A ``large'' bare charge transfer energy, $\Delta_{CT}\sum_{i}(n_{i,b}-n_{i,a})=\Delta_{CF}\sum_{i}[(1-n_{i,a})-(1-n_{i,b})]=\Delta_{CT}\sum_{i}(P_{i,a}^{z}-P_{i,b}^{z})$ already exists in DFT or tight-binding calculations (see Fig.~\ref{figS1} of SI).  Adding this term to $H_{RKKY}^{zz}$
 results in an orbital polarization renormalized by $H_{res}$ (there is a separate renormalization of the orbital occupations and thus $\Delta_{CT}$ from DMFT, see main text).  Thus, the $zz$-ordered phase of this model in a ``large Zeeman field'' gives ferro-orbital order, with $\langle P_{i,\mu}^{z}\rangle = (1-\langle n_{i,\mu}\rangle)\neq 0$ ($\mu=a,b$) with $\langle n_{i,a}\rangle > \langle n_{i,b}\rangle$.  Goodenough-Kanamori-Anderson rules then mandate antiferromagnetic inter-site order or spin correlations.  Moreover, a $d$-wave bond-excitonic order, resulting as a condensate of inter-band excitons, emerges as a leading competitor to $d$-wave SC: this is already clearly encapsulated in the very form of $H_{res}^{(exc)}$.  Thus, both $d$-wave SC and ``ferro-orbital''-plus-$d$-wave bond-excitonic orders arise as direct instabilities of an OSMP or FFL state.  Even with small spin-orbit coupling, there will be an admixture of triplet excitonic component, making it possible for the circulating current (CC) phase to appear as a ``competitor'' to SC.

\vspace{0.5cm}
 
{\bf SI3: Reconciliation with Quasi-One Dimensional Magnetism in BCO} In BCO, DFT and tight-binding calculations yield a one-dimensional ($D=1$) $d_{z^{2}}$ band 
hybridized with a $D=2$
band of $d_{x^{2}-y^{2}}$ band via a $d$-wave form-factor hybridization.  Upon adding electronic correlations, one might expect this one-dimensionality to get more pronounced,
leading to possible Luttinger liquid behavior.  Indeed, recent data~\cite{s7Li2025} reveal the good accord with both, the two-spinon continuum and the des Cloizeaux-Pearson  dispersion, both characteristic of a $D=1$ Heisenberg spin chain!  This is associated with the unpaired electron in the $d_{z^{2}}$ orbital in undoped BCO, where direct hopping between
neighboring $d_{z^{2}}$ orbitals on CuO chains can lead to such an outcome.  The very observation of a two-spinon continuum and the des Cloizeaux-Pearson ``spin-wave'' dispersion, characteristic of the $D=1$ spin-$1/2$ Heisenberg model, is a conclusive argument in favor of strong electronic correlations and, in doped BCO, to an orbital-selective metal.  It contradicts a non- or weakly interacting view, where no such features can obtain due to absence of an OSMP.  In any DFT or weak-coupling framework, the $d$-wave hybridization with $d_{x^{2}-y^{2}}$ states is always relevant, and must drive 
BCO toward an anisotropic $D=3$ Landau FL metal.  Upon doping BCO ($x\neq 0$), this tendency can only get enhanced.  This would be in direct conflict with observation of $D=1$ Heisenberg model spin fluctuations in RIXS data in BCO.

In a Mott-Hubbard picture, however, things are drastically
modified: the lone hole in the Mott localized $d_{z^{2}}$ orbital will readily yield a spin-$1/2$ $D=1$ Heisenberg spin chain due to Anderson super-exchange along the Cu-O
link.  If these states remain in Mott insulating states upon doping BCO, as we find in the main text, such $D=1$ Luttinger liquid behavior of the spin chain will persist in doped BCO as well.
While this {\it is} seen, the accord with data is much better for $x=0$.  For the doped case, the accord is less satisfying.  It is found that supplementing the physics of the Heisenberg spin-$1/2$ chain by an additional {\it attractive} interaction leads to more satisfying agreement in the latter case.  

In our OSMP picture, we can rationalize these findings as follows: for $x=0$, since the Mott localized $d_{z^{2}}$ states with a single electron per site sits at higher energy,
direct Cu-Cu super-exchange directly leads to a $S=1/2$ Heisenberg chain, in excellent accord with data.  Upon doping ($x>0$), we invoke our OSMP picture, along with the 
residual interaction derived in the main text.  The selectively localized $d_{z ^{2}}$-fermions still contribute a $D=1$ spin-$1/2$ Heisenberg chain as before.  But we now 
have an additional interaction $H_{res}$.  To see its effect, we first use the identity $({\bf S}_{i}.{\bf S}_{j}-1/4)=-b_{ij}^{\dag}b_{ij}$ with $b_{ij}^{\dag}=\sum_{\sigma}\sigma b_{i\sigma}^{\dag}b_{j,-\sigma}^{\dag}/\sqrt{2}$ along with the form of $H_{res}$ to obtain the model

\be
H_{1}=-J_{c}\sum_{<i,j>}b_{ij}^{\dag}b_{ij} -\sqrt{2}J_{ab}\sum_{<i,j>}(\sigma a_{i\sigma}^{\dag}a_{j,-\sigma}^{\dag}b_{ij}+h.c) + H_{a}
\ee
where $J_{c}\simeq 4t_{bb}^{2}/U$.  $H_{a}$ describes the $a$-fermion sector.  But the first two terms represent precisely the system of $D=1$ spin-$1/2$ chains subject to an attractive pairing potential that arises from $H_{res}$.  It transfers a ``preformed'' singlet pair from the $b$-fermion sector (the $D=1$ CuO chains) to the $a$-fermion sector.  This is thus not inconsistent with the additional attractive potential necessary to achieve accord with data~\cite{s7Li2025} in doped BCO.  Indeed, the attractive nearest-neighbor interaction along the chain, $H_{V}=-V\sum_{<i,j>,\sigma,\sigma'} n_{ib\sigma}n_{jb\sigma'}=-V\sum_{<i,j>}[b_{i\uparrow}^{\dag}b_{j\downarrow}^{\dag}b_{j\downarrow}b_{i\uparrow} + n_{ib\uparrow}n_{jb\uparrow}
+ (\uparrow \rightarrow \downarrow)]$.  The first term can be generated by integrating out the $a$-fermion sector in $H_{1}$ above.  Alternatively, we notice that, away from the half-filled chain limit, we have $({\bf S}_{i}.{\bf S}_{j}-n_{ib}n_{jb}/4)=-b_{ij}^{\dag}b_{ij}$, again consistent with an additional n.n attractive interaction between $b$-holes.
  Thus, our OSMP picture can lead to a model similar to that used in the RIXS experiment, even without invoking coupling to phonons.  Of course, this coupling could be added to 
  the above: it would provide an additional attractive interaction between $b$-fermions, but this cannot affect our arguments above.
  
In the alternative case where SC from $b$-band pairs could arise, this would also necessarily have to follow a non-standard pairing mechanism.  Though we do not find SC in this sector in this paper, we discuss this case for completeness. 
On a theoretical level, observation of Luttinger liquid behavior may suggest that a picture related to that used for deriving SC instability from coupled Hubbard chains~\cite{s8Biermann} could be operative.  Recollect that our $H_{res}$ hops a Cooper pair from the $b$-sector (the $D=1$ Heisenberg chain in the OSMP, as above) to the $a$-fermion band and vice-versa.  One may view this as follows:  in coupled Hubbard chains, an irrelevant interchain one-fermion hopping, $t_{\perp}$, nonetheless leads to relevant two-fermion hopping processes to $O(t_{\perp}^{2})$.  It results in a direct instability of the Luttinger liquid to either density-wave or SC orders.  In our case, this role is played by $V_{ab}(k)$:  it is irrelevant to $O(V_{ab})$ in the OSMP, but generates $H_{res}$ at $O(V_{ab}^{2})$.  A look at it's structure reveals that the spin-singlet (or two-spinon) pair on the $D=1$ chains can hop off (via $H_{res}$) into the $a$-band, and again hop off into a neighboring spin chain via the same process in reverse.  The inter-band proximity effect will, of course, also induce SC in the $a$-band, leading to a multi-band SC.  It maybe possible to develop an analytic framework for such a mechanism, using bosonization techniques.  In spite of this 
resemblance to coupled Luttinger chains, however, the situation in our model is more complex: instead of a single parameter, $t_{\perp}$, the Luttinger chains are coupled via 
a two-dimensional band of correlated $a$-fermion states.  Such an alternative requires a non-trivial extension of the picture of coupled Luttinger liquids.  It corresponds to a distinct pair state with different pair-symmetry: we have not studied this case here.
    
Experimentally, observation of two-spinon dispersion characteristic of a spin-$1/2$ $D=1$ Heisenberg chain is a crucial point.  It can be viewed as a ``smoking gun'' to distinguish our proposal from the alternatives, where Cooper pairing, mediated by $D=1$ spin fluctuations in the $b$-band sector, arises in the $b=d_{z^{2}}$ band.  If this were the case, the gapless two-spinon feature seen in RIXS data would have to reveal a spin gap upon a transition to the SC phase.  Persistence of the gapless des-Cloizeaux Pearson-like spin-wave spectrum even in the doped sample~\cite{s7Li2025} is thus, in our opinion, strong evidence against Cooper pairing in this sector.

  We should mention that SC in systems with co-existent $D=1$ and $D=2$ bands is not entirely new.  It is long known that such a situation obtains in Sr$_{2}$RuO$_{4}$ (SRuO) which, 
  however, shows a novel SC below $T_{c}=1.5$~K, without doping.  There, the inter-band hybridization couples two $D=1$ $d_{xz,yz}$ bands and a single $D=2$ $d_{xy}$ band states.
  In BCO, a $D=1$ $d_{z^{2}}$ band mixes with the $D=2$ $d_{x^{2}-y^{2}}$ band.  A low $T$ coherence-incoherence crossover is also seen in SRuO, and attributed to either Hund metallicity or to Luttinger liquids perturbed by mixing with a $D=2$ band~\cite{s9Acharya2017}.  At low $T$, a relevant hybridization in SRuO drives it to a correlated Landau FL fixed point, and SC arises from a metal with well-defined low energy Landau quasiparticles.  In BCO, in sharp contrast, the SC $T_{c}=73$~K is larger than the scale at which this crossover occurs, as we have shown in the main text, so SC emerges directly from an incoherent, strange-like metal.  Since an OSMP produces partially unquenched local moments, similar to Hund's metal views, the Hund metal or OSMP-versus-Luttinger liquid views are not incompatible: once local moments form in the OSMP, direct or indirect (via the metallic $a$-fermion states) super-exchange will lead to $D=1$ spin chains and Luttinger liquid physics perturbed by coupling to the $D=2$ dimensional band of $d_{x^{2}-y^{2}}$ states.    
  
\vspace{0.5cm}

{\bf SI4: Role Of Interlayer Hopping And $d$-wave SC in BCO}:  What about interlayer coupling in BCO?  This should drive the full three-dimensional SC.  In canonical cuprates, the form factor of the interlayer hopping, $t_{\perp}(k)\simeq ($cos$k_{x}$-cos$k_{y})^{2}$, perfectly facilitates the interlayer hopping of in-layer $d$-wave pairs due to ``nodal protection'': we call this an ``orbital form-factor matching''.  This feature is much more complex in BCO in BCO, because of its multi-orbital band structure.  For the $b=d_{z^{2}}$ orbital, the BCO structure constrains the interlayer hopping to have the form

\be
t_{\perp}(k)=t_{\perp} cos(k_{x}/2)cos(k_{y}/2)cos(k_{z}/2)
\ee
This will give rise to a ``$c$-axis'' counterpart of the residual interaction, given by
$1, cosk_{x}, (cosk_{x}+cosk_{y})cosk_{z}, cosk_{x}cosk_{y}cosk_{z}, cosk_{x}cosk_{y}$, etc, {\it i.e}, a combination of $s$, $s_{\pm},p$ and $d_{xy}$ symmetries, among 
others.  Hence, irrespective of the pair symmetry of inlayer pairs, there is no such ``matching'' between inlayer and interlayer pair form factors in BCO.  Such situations are not uncommon in multi-orbital transition-metal oxides:  their role in high-T$_{c}$ SC in nickelates, Fe arsenides and Sr$_{2}$RuO$_{4}$ has been studied by various groups~\cite{s10Schnyder}, and this mismatch of in- and inter-layer pair form factors can trigger phase competition and appearance of additional bond-order wave phase(s) competing with SC.  The $p$-wave component must interfere destructively with the $s,s_{\pm},d_{xy}$ pair channels.  Furthermore, it must also be much more sensitive to doping-induced disorder.  In general, these effects will negatively affect the possibility of a three-dimensional $T_{c}$.  

But the interlayer hopping form factor for the hopping of $d_{x^{2}-y^{2}}$ pairs has a much more favorable form.  Specifically,

\be
\begin{split}
H_{\text{res}}^{\perp\prime} \simeq {} & \left( \frac{4t_{\perp}^{\prime 2}}{U'} \right) \sum_{k,k',\mu,\mu',\sigma,\sigma'} \lambda(k)\lambda(k') (a_{k,\mu,\sigma}^{\dag}a_{k,\mu',\sigma} + \text{h.c.}) \\
& (a_{k',\mu',\sigma'}^{\dag}a_{k',\mu,\sigma'} + \text{h.c.})
\end{split}
\ee
with $\lambda(k)=($cos$k_{x}-$cos$k_{y})^{2}$cos$(k_{z}/2)$.  This form factor is much more optimally ``orbitally matched'' for tunnelling of $a$-band Cooper pairs between neighboring layers and, thus, to induce full three-dimensional SC.  Given the planar $d$-wave pairing derived in the main text, it follows that this interlayer form-factor 
will lead to a full $D=3$ $d$-wave superconductor.  On the other hand the strong orbital mismatch for $d_{z^{2}}$-pair hopping will significantly frustrate $D=3$ SC in that sector.  We propose that $d_{x^{2}-y^{2}}$-wave SC wins, and co-exists with perturbed LL physics in the $d_{z^{2}}$ sector.  Thus, a normal-state OSMP undergoes a direct instability to a {\it fractionalized superconductor}.

\end{document}